\documentclass[sigconf]{acmart}
\usepackage{multicol}
\usepackage{booktabs}
\usepackage{subcaption}
\usepackage{enumitem}[itemindent=1em]

\usepackage{bm}
\usepackage{balance}

\newtheorem{definition}{Definition}

\usepackage{float}

\usepackage[linesnumbered,ruled,vlined]{algorithm2e}

\AtBeginDocument{%
  \providecommand\BibTeX{{%
    \normalfont B\kern-0.5em{\scshape i\kern-0.25em b}\kern-0.8em\TeX}}}
\newcommand{\model}{PS2}

\copyrightyear{2023}
\acmYear{2023}
\setcopyright{acmcopyright}\acmConference[WSDM '23]{Proceedings of the Sixteenth ACM International Conference on Web Search and Data Mining}{February 27-March 3, 2023}{Singapore, Singapore}
\acmBooktitle{Proceedings of the Sixteenth ACM International Conference on Web Search and Data Mining (WSDM '23), February 27-March 3, 2023, Singapore, Singapore}
\acmPrice{15.00}
\acmDOI{10.1145/3539597.3570407}
\acmISBN{978-1-4503-9407-9/23/02}

\begin{document}
\title{Bring Your Own View: Graph Neural Networks for Link Prediction with Personalized Subgraph Selection}




\author{Qiaoyu Tan}
\affiliation{
  \institution{Texas A\&M University}
\country{}
  }
\email{qytan@tamu.edu}

\author{Xin Zhang}
\affiliation{
  \institution{The Hong Kong Polytechnic University}
 \country{Hong Kong SAR}
  }
\email{xin12.zhang@connect.polyu.hk}

\author{Ninghao Liu}
\affiliation{
  \institution{University of Georgia}
 \country{USA}
  }
\email{ninghao.liu@uga.edu}

\author{Daochen Zha}
\affiliation{
  \institution{Rice University}
 \country{USA}
  }
\email{daochen.zha@rice.edu}

\author{Li Li}
\affiliation{
  \institution{Samsung Electronics America}
\country{USA}
  }
\email{li.li1@samsung.com}

\author{Rui Chen}
\affiliation{
  \institution{Samsung Electronics America}
 \country{USA}
  }
\email{rui.chen1@samsung.com}

\author{Soo-Hyun Choi}
\authornote{Corresponding author.}
\affiliation{
  \institution{Samsung Electronics}
\country{USA}
  }
\email{sh9.choi@samsung.com}

\author{Xia Hu}
\affiliation{
  \institution{Rice University}
 \country{USA}
  }
\email{xia.hu@rice.edu}
\renewcommand{\shortauthors}{Q. Tan, et al.}

\begin{abstract}
Graph neural networks (GNNs) have received remarkable success in link prediction (GNNLP) tasks. Existing efforts first predefine the subgraph for the whole dataset and then apply GNNs to encode edge representations by leveraging the neighborhood structure induced by the fixed subgraph. The prominence of GNNLP methods significantly relies on the adhoc subgraph. Since node connectivity in real-world graphs is complex, one shared subgraph is limited for all edges. Thus, the choices of subgraphs should be personalized to different edges. However, performing personalized subgraph selection is nontrivial since the potential selection space grows exponentially to the scale of edges. Besides, the inference edges are not available during training in link prediction scenarios, so the selection process needs to be inductive. To bridge the gap, we introduce a Personalized Subgraph Selector (\model) as a plug-and-play framework to automatically, personally, and inductively identify optimal subgraphs for different edges when performing GNNLP. \model~is instantiated as a bi-level optimization problem that can be efficiently solved differently. Coupling GNNLP models with \model, we suggest a brand-new angle towards GNNLP training: by first identifying the optimal subgraphs for edges; and then focusing on training the inference model by using the sampled subgraphs. Comprehensive experiments endorse the effectiveness of our proposed method across various GNNLP backbones (GCN, GraphSage, NGCF, LightGCN, and SEAL) and diverse benchmarks (Planetoid, OGB, and Recommendation datasets). Our code is publicly available at \url{https://github.com/qiaoyu-tan/PS2}
  
\end{abstract}




\maketitle


\section{Introduction}
Graph is a ubiquitous and powerful data structure to present different types of relational data, such as social networks and biological molecules. Given that real-world graphs are often only partially observed, \textit{link prediction}~\cite{zhou2021progresses}, which aims to predict missing links in a graph, is a central problem across many scientific domains. For example, link prediction has applications in predicting protein interactions~\cite{qi2006evaluation}, drug responses~\cite{stanfield2017drug}, and completing the knowledge graph~\cite{arora2020survey,dong2023active}. Besides, it is also the backbone for various recommendation systems, e.g., friend suggestion in social networks~\cite{adamic2003friends,tan2019deep} or product recommendation in online market-places~\cite{ying2018graph,tan2021dynamic,zha2022dreamshard}. 

Recently, considerable efforts have been made to develop advanced link prediction techniques~\cite{gao2021graph,cai2021line,yang2019homogeneous}. Among them, graph neural networks (GNNs) based link prediction models (GNNLP) have achieved impressive results~\cite{kipf2016variational,hamilton2017inductive,wang2019neural,he2020lightgcn,zhang2018link,zhang2021labeling,tan2020learning}, owing to the expressive encoding capacity of GNNs. 
The essential idea behind GNNLP is to generate edge representation based on the subgraph around the anchor edge via a GNN encoder and then estimate its likelihood with a prediction function. According to the difference in utilizing subgraph for edge embedding, they can be divided into two categories: node2link~\cite{kipf2016variational,hamilton2017inductive,he2020lightgcn} and subgraph2link~\cite{zhang2018link,zhang2021labeling,pan2021neural}. The node2link approaches (i.e., GAE~\cite{kipf2016variational}, GraphSage~\cite{hamilton2017inductive}, and LightGCN~\cite{he2020lightgcn}) aim to first learn node representations for the head and tail nodes of the anchor edge independently, and then combine the representations of end nodes for edge embedding. In contrast, the subgraph2link approaches (e.g., SEAL~\cite{zhang2018link} and~\cite{zhang2021labeling}) target to learn edge representation by pooling over the subgraph of the given edge, casting it as a graph representation learning task.

While effective, both of them have largely overlooked the diversity of subgraphs when embedding different edges.
For example, GAE and SEAL assume that the best subgraph structure for all edges is the same and adopt the neighbors within $k$-hops for edge embedding, where $k\in\{1,2,\cdots,K\}$ is a hyperparameter. Although such collective selection can significantly reduce the tedious tuning efforts to identify the best $k$ value from $K$ options, the shared subgraph structure assumption is rather limited. Different edges may favor different subgraph structures for link prediction. This hypothesis is reasonable because node connectivity patterns in real-world graphs are complex~\cite{lu2011link,zhang2022graph}. For instance, in social networks, the social connectivity of users is created by different factors~\cite{aiello2012link,liu2019single}. In the recommendation system, a user's purchase behavior could be motivated by either his/her like-minded customers or conceptually similar products in the historical records~\cite{tan2021sparse}. 

\begin{figure}[htp]
\vspace{-0.4cm}
    \centering
    \includegraphics[width=6.8cm]{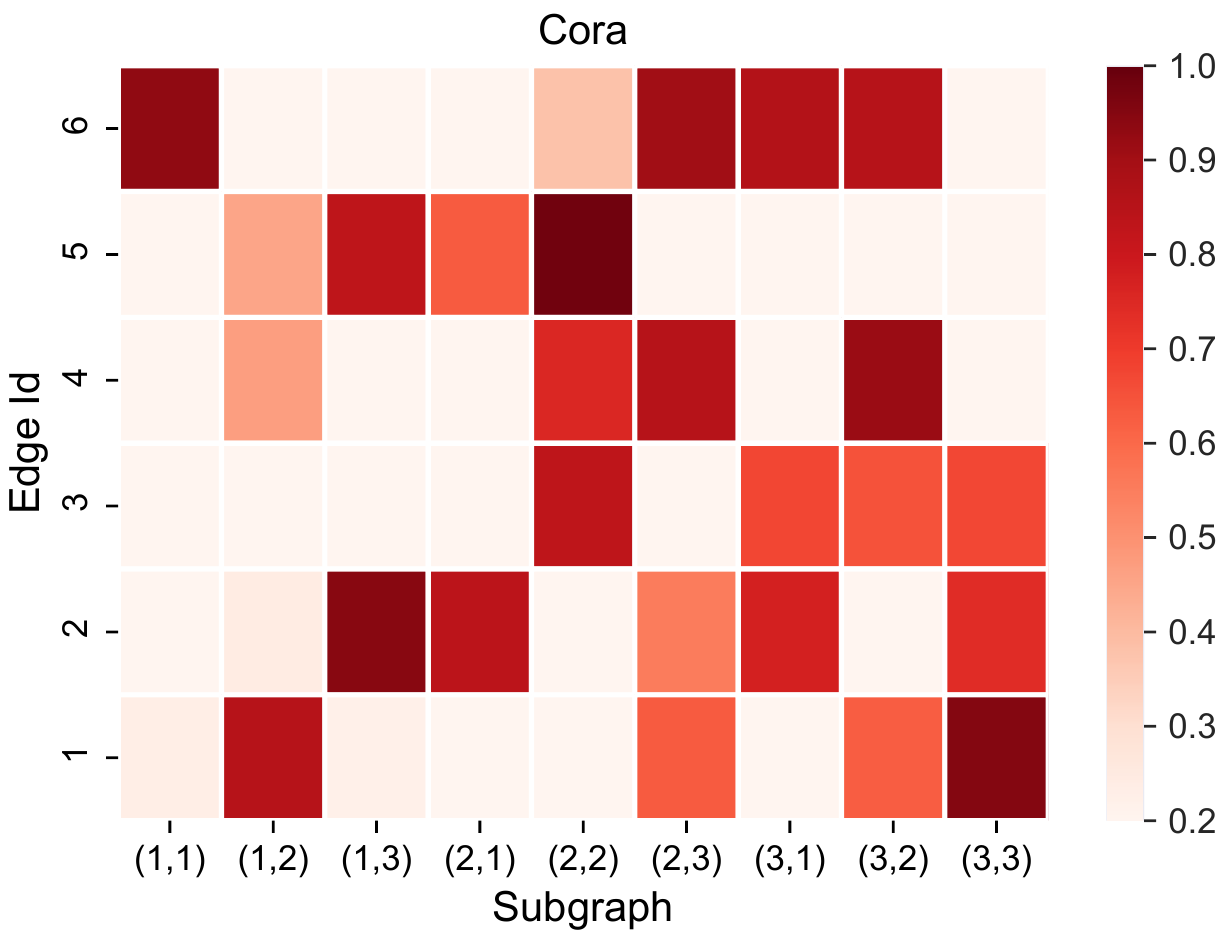}
    \caption{The effect of different edge subgraphs towards 6 randomly sampled test edges on the Cora dataset. The results are obtained by training GAE~\cite{kipf2016variational} nine times with different subgraph ranges. For example, $(2,3)$ means the subgraph of an edge is composed of the 2-hops and 3-hops of neighbors of its head and tail nodes, respectively. The X-axis denotes different subgraphs selection strategies, and the Y-axis is the id of the sampled edges. The color from light to dark represents the predicted probability for edge existence. }
    \label{fig:observation}
    \vspace{-0.4cm}
\end{figure}

Motivated by this, we conduct a preliminary experiment on the Cora dataset to test how different subgraph structures impact the link prediction results in Figure~\ref{fig:observation}. We observe that GAE can accurately infer different missing edges by training over various predefined subgraph structures. For instance, the first missing edge can be well recovered by using the neighbors within 3 hops, while the second missing edge can be effectively reconstructed by using the 2-hops and 3-hops of neighbors of its head and tail nodes, respectively. In terms of their input neighborhood subgraphs, this personalized phenomenon of edges, has never been explored in link prediction scenarios. To bridge the gap, in this paper, we propose to develop an effective subgraph selector to automatically identify the most informative subgraphs for different edges.

However, it is a nontrivial and challenging task mainly because of three roadblocks. First, given a graph data, the latent subgraph selection space is exponential to the size of edges, which is millions or even billions in practice. It is impossible to identify the optimal subgraph configurations for all edges via the brute-force search. Second, in link prediction applications, the edges to be predicted are not available during the training. Thus, the subgraph selection process must be inductive, enabling infer subgraph structures for unseen edges. Third, how to make the edge-wise subgraph selection adaptive to the well-established base models such as methods under the node2link approach (e.g., GAE~\cite{kipf2016variational}, GraphSage~\cite{hamilton2017inductive}, NGCF~\cite{wang2019neural} and LightGCN~\cite{he2020lightgcn}) or subgraph2link category (e.g., SEAL~\cite{zhang2018link}). 

To address these challenges, we propose a novel personalized subgraph selector, dubbed \model, as a plug-and-play framework. It aims to develop an automatic and inductive subgraph selection module for GNNLP methods, such that the most informative subgraph structures can be explicitly identified and exploited for different edges. Specifically, we aim to explore two important research questions. (i) How to automatically sample the optimal subgraph structure for each edge efficiently, and make the selection process inductive? (ii) How to effectively equip well-established GNNLP methods with the proposed personalized selector, so as to offer orthogonal gains across a variety of graph domains and GNNLP backbones? We summarize our major contributions as follows. 

\begin{itemize}[leftmargin=*]
    \item We focus on subgraph selection for GNNs based link prediction (GNNLP) problem, and propose an effective personalized subgraph selector (\model). \model~is the first to automate subgraph selection in an edge-wise fashion when performing GNLLP. It can be easily adopted to boost the well-studied GNNLP methods. 
    \item \model~can be formulated under bi-level optimization and solved using the alternating gradient-descent algorithm. It is inspired by the differentiable architecture search~\cite{liu2018darts}, but we extend it from a transductive search model to an inductive subgraph selector, focusing on the edge-wise subgraph structure selection rather than the model architecture search.
    \item We conduct extensive experiments to evaluate \model~on multiple graph benchmarks of diverse types and scales, over a variety of GNNLP backbones. Empirical results show that with \model, the performance of state-of-the-art GNNLP competitors can be advanced with a wide margin.  
\end{itemize}

\begin{figure*}[t]
\vspace{-0.4cm}
\begin{center}
\includegraphics[width=16.8 cm]{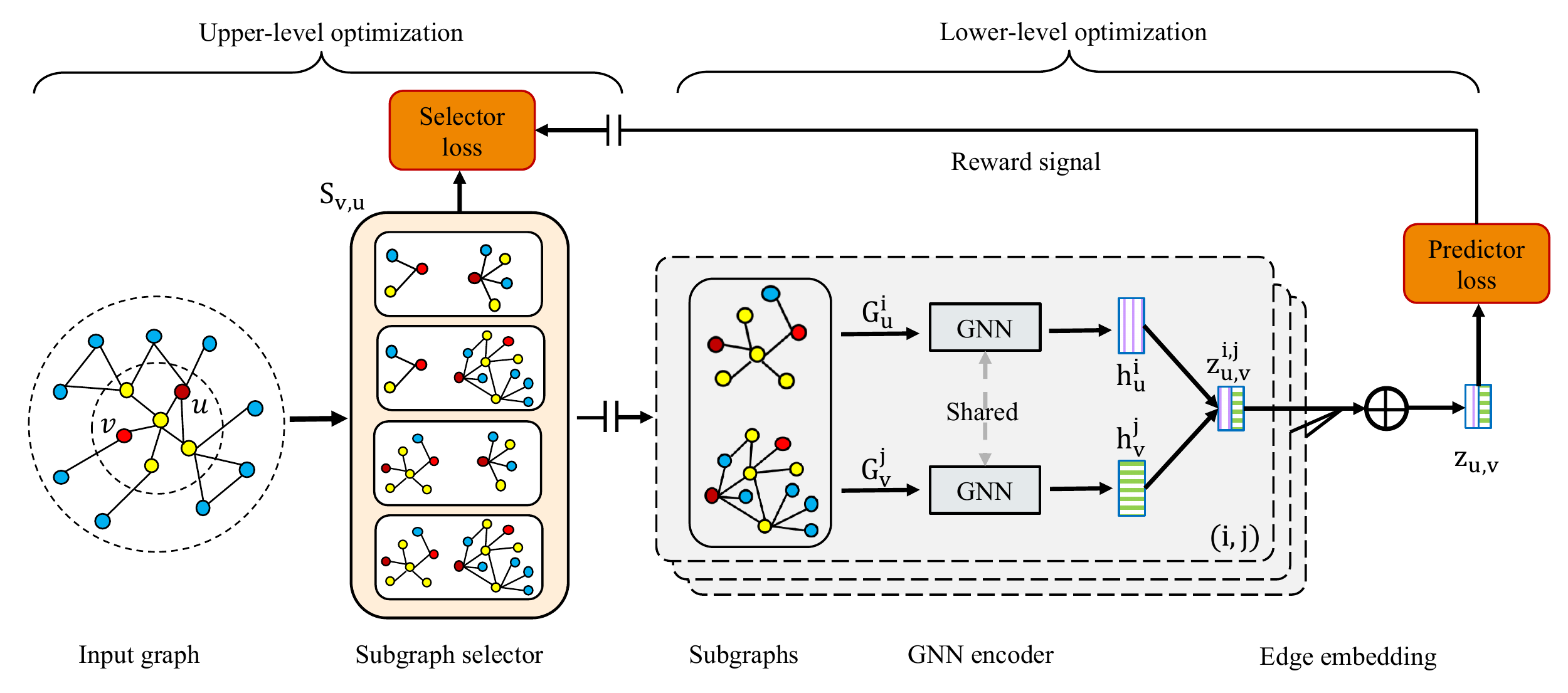}
\caption{{The training pipeline (i.e., search phase) of the proposed \model. The subgraph selector generates an importance score for each subgraph candidate. The upper-level optimization updates the parameters of the selector $g_{\bm{\theta}}$. The lower-level optimization updates the parameters of the GNN $f_{w}$ and predictor $q_{w}$.}}
\label{fig:mgae}
\end{center}
\vspace{-0.5cm}
\end{figure*}

\section{Preliminaries}
\textbf{Problem Formulation.} We are given a graph $\mathcal{G}=(\mathcal{V}, \mathcal{E})$ with $n$ nodes, where $\mathcal{V}$ and $\mathcal{E}$ denote the sets of nodes and edges, respectively.  We use $\mathcal{G}_v^k$ to represent the local subgraph of node $v\in\mathcal{V}$, and $\mathcal{G}_{u,v}^k=(\mathcal{G}_u^k, \mathcal{G}_v^k)$ denotes the subgraph of edge $e_{u,v}\in\mathcal{E}$ within $k$ hops, in which $k$ is a crucial hyperparameter. Note that $\mathcal{G}_{u,v}^k$ is obtained by removing the redundancies (e.g., repeated edges and nodes) in subgraphs $\mathcal{G}_u^k$ and $\mathcal{G}_v^k$. In previous studies, the optimal $k$ value is manually picked from the set $\{1,2,\cdots, K\}$ via either rule of thumb or validation, where the same $k$ is applied to the whole graph $\mathcal{G}$. 
However, as shown in Figure~\ref{fig:observation}, the optimal $k$ varies for predicting different edges. Therefore, we propose a personalized subgraph selection to identify the most informative subgraph for different edges, where the problem is formally defined below.  

\begin{definition}
\textbf{Personalized subgraph selection.} Given a graph $\mathcal{G}=(\mathcal{V}, \mathcal{E})$, the subgraph space for node $v\in\mathcal{V}$ is defined as $\mathcal{S}_v=\{\mathcal{G}^1_v, \mathcal{G}_v^2, \cdots, \mathcal{G}_v^k, \cdots, \mathcal{G}_v^K\}$. 
To predict edge $e_{u,v}$, the personalized subgraph selection aims to find the optimal subgraph $\mathcal{G}_{u,v}=(\mathcal{G}_u^i, \mathcal{G}_v^j | \mathcal{G}_u^i\in\mathcal{S}_u, \mathcal{G}_v^j\in\mathcal{S}_v)$. The values of $i$ and $j$ are determined adaptively for different target edges.
\label{def:PS2}
\end{definition}

Compared with existing GNNLP methods, our personalized setting implies two essential properties as below. 
 
\begin{itemize}[leftmargin=*]
    \item \textbf{Edge subgraph is personal.} In our setting, the local subgraphs are personalized to different edges. For example, the subgraph can be $\mathcal{G}_{1,2}=(\mathcal{G}_1^1, \mathcal{G}_2^3)$ in predicting edge $e_{1,2}$, but it also can be $\mathcal{G}_{3,4}=(\mathcal{G}_3^2, \mathcal{G}_4^1)$ for edge $e_{3,4}$. However, in existing GNNLP efforts, the subgraph order is restricted to be the same for all edges, i.e., $\mathcal{G}_{1,2}=(\mathcal{G}_{1}^2, \mathcal{G}_2^2)$ and $\mathcal{G}_{3,4}=(\mathcal{G}_3^2, \mathcal{G}_4^2)$.
    \item \textbf{Node subgraph is polysemous.} Node $v$ can use different subgraphs in predicting different edges. For example, the optimal subgraph for $e_{1,2}$ is $\mathcal{G}_{1,2}=(\mathcal{G}_1^1, \mathcal{G}_2^3)$, while $\mathcal{G}_{1,4}=(\mathcal{G}_1^3, \mathcal{G}_4^2)$ for edge $e_{1,4}$. In existing GNNLP methods, the neighbor range of node subgraph is fixed, e.g., subgraphs for node $1$, $2$, and $4$ are $\mathcal{G}_1^2$, $\mathcal{G}_2^2$, and $\mathcal{G}_4^2$, respectively. 
\end{itemize}

\noindent\textbf{GNNs for Node Embedding.} Given a node $v\in\mathcal{V}$, GNNs models~\cite{kipf2016semi,gilmer2017neural} are widely adapted to mapping nodes into hidden representations, i.e., $f_w: \mathcal{V}\rightarrow \mathbb{R}^d$. GNNs target to update the node presentation by aggregating representations of itself and its neighbors. Formally, at the $k$-th layer, we have 
\begin{equation}
\mathbf{h}^{(k)}_v=\text{UPDATE}(\mathbf{h}^{(k-1)}_v, \text{AGGREGATE}(\{\mathbf{h}_{u}^{(k-1)}: u\in\mathcal{N}_v\})).
\label{eq:mps}
\end{equation}
$\mathbf{h}_v^k\in\mathbb{R}^d$ is the hidden representation of node $v$ at the $k$-th layer, while $\mathcal{N}_v$ is the set of nodes adjacent to $v$. We often initialize $\mathbf{h}_v^{0}$ as $\mathbf{x}_v$.
The AGGREGATE function aims to receive messages from neighbors and the UPDATE function focuses on updating $v$'s representation based on the representation from the previous GNN layer and the information from neighbors. By stacking $K$ GNN layers, each node $v$ has $K$ hidden representations $\{\mathbf{h}_v^1, \cdots, \mathbf{h}_v^K\}$.



\section{The Proposed Method}
In this section, we present the details of the proposed \model~shown in Figure~\ref{fig:mgae}.
We first discuss the exponential subgraph selection space of our problem. Then, we elaborate on a tailored inductive subgraph selector to effectively sample subgraphs for seen and unseen edges in this space. Finally, we show how to formulate our training objectives into bi-level optimization and solve it via alternating gradient descent. 

\subsection{Subgraph Selection Space}
Given a graph $\mathcal{G}=(\mathcal{V}, \mathcal{E})$ and the maximum number of hops $K$ considered, there are $K$ latent subgraphs for each node $v\in\mathcal{V}$, denoted by $\mathcal{S}_v=\{\mathcal{G}_v^k\}_{k=1}^K$. Each subgraph $\mathcal{G}_v^k$ is spanned by the anchor node $v$ and its neighbors within $k$ hops, whose shortest path distance from $v$ is less than $k$. In previous studies, $k$ is a dataset-level hyperparameter, which is fixed for all nodes/edges in the graph $\mathcal{G}$. 
In this work, we denote the subgraph of edge $e_{u,v}$ as $\mathcal{G}_{u,v}^k=(\mathcal{G}_u^k, \mathcal{G}_v^k)$. In practice, the best $k$ value is usually selected through validation and is applied to the whole graph. This collective selection strategy has been widely adopted as the default protocol in prior GNNLPs. 

However, as discussed before, applying a constant $k$ value in selecting subgraphs leads to suboptimal results for predicting some edges. Thus, we propose to adaptively choose the subgraph selection space for different edges. Specifically, we define the subgraph of edge $e_{u,v}$ to be composed of its end nodes' subgraphs. That is, $\mathcal{G}_{u,v}=(\mathcal{G}_u^i, \mathcal{G}_v^j| 1 \leq i, j \leq K)$. Then, the potential subgraph pool size for each edge is $K^2$, and {the total subgraph selection space equals to ${K}^{2|\mathcal{E}|}$ for the whole graph, where $|\mathcal{E}|$ is the number of edges in $\mathcal{G}$.} Although $K$ is empirically small (e.g., $K=2, 3$) in link prediction scenarios, the selection space in our personalized setting is still huge and intractable as \textbf{the complexity grows exponentially with the edge size}. For example, when $K=2$ and $|\mathcal{E}|=50$, we have nearly $10^{30}$ selection candidates. The situation is more difficult in real-world graphs, where $|\mathcal{E}|$ is millions or even billions. 

In summary, by personalizing edge subgraphs, the subgraph selection space for link prediction increases from $K$ to $K^{2|\mathcal{E}|}$. Therefore, existing strategies based on the rule of thumb or grid search are no longer appropriate. Also, a tailored subgraph space selector is needed to tackle our personalized subgraph selection problem. 



\subsection{Personalized Subgraph Selector}
\label{sub:subselector}
To assign different subgraph orders to different edges for link prediction, the intuitive solution is random selection. For example, given an edge $e_{u,v}$, we can randomly select its subgraph (e.g., $\mathcal{G}_{u,v}=(\mathcal{G}_u^1, \mathcal{G}_v^3)$) from the candidate pool. Despite the simplicity, the random selection approach fails to control the quality of the resulting subgraphs. Coupling existing GNNLP methods with random subgraph selection could incur significant performance degradation, especially when the graph is challenging, e.g., on OGB datasets (see Table~\ref{table_lp_ogb}).

To address this issue, we focus on data-driven selection by making the subgraph selection process learnable. The core idea is to parameterize the subgraph selector with a deep neural network, which takes a query edge $e_{u,v}$ as input and outputs its optimal subgraph for downstream link prediction. 
The main challenges to achieving this goal are two-fold. (i) Given the exponential complexity (i.e., $K^{|\mathcal{E}|}$) of the selection space, how can we make the personalized subgraph selection scale to real-world graphs with millions or even billions of edges? (ii) Since the edges to be inferred are not available in training under link prediction scenarios, how to make the subgraph selector inductive to unseen edges? We introduce our solutions below. 

\subsubsection{Learnable Subgraph Selector.} To tackle the first challenge, we propose to make the selection process learnable. Given an edge $e_{u,v}$ and its $K^2$ subgraph candidates, i.e., $\{(\mathcal{G}_u^i, \mathcal{G}_v^j)\}_{i,j=1}^K$, our subgraph selector aims to find the most informative subgraph from the candidate set. This selection problem is well-known to be discrete and non-differentiable. While enormous efforts based on evolution~\cite{real2019regularized} or reinforcement learning~\cite{zoph2018learning} have been dedicated to addressing the discrete selection problem, they are inefficient for training. To tackle this problem, we make the selection process learnable by relaxing the discrete selection space to be continuous. The core idea is to relax the selection of a single subgraph to a softmax over all possible subgraph candidates. Formally, let $\alpha_{u,v}^{i,j}$ denote the contribution of subgraph $(\mathcal{G}_{u}^i, \mathcal{G}_v^j)$ in predicting edge $e_{u,v}$, the learnable selection process is defined as:
\begin{equation}
\begin{split}
    \mathcal{G}_{u,v}=\sum_{1\leq{i,j}\leq K}\frac{\exp (\alpha_{u,v}^{i,j}/\tau)}{\sum_{i',j'=1}^K \exp(\alpha_{u,v}^{i',j'}/\tau)} * \mathcal{G}_{u,v}^{i,j},
\end{split}
\label{eq:learnable}
\end{equation}
we use $*$ to denote the multiplication alike operation on the subgraph in theory. $\alpha_{u,v}=[\alpha_{u,v}^{1,1}, \cdots, \alpha_{u,v}^{K,K}]\in\mathbb{R}^{K^2}$ is the subgraph related weight vector for edge $e_{u,v}$, which is initialized as part of model parameters. $\tau$ is a small temperature parameter, which helps approximate the categorical selection distribution. By Eq.~\eqref{eq:learnable}, the subgraph selection process reduces to learning a set of continuous variables $\{\bm{\alpha}_{u,v}: e_{u,v}\in\mathcal{E}\}$. After the subgraph selector is well-trained, a discrete subgraph selection can be acquired by replacing the mixed selection with the most likely subgraph, i.e., $\mathcal{G}_{u,v}=\mathcal{G}_{u,v}^{i,j}$ if $\alpha_{u,v}^{i,j}=\arg\max_{i',j'} \alpha_{u,v}^{i',j'}$.

Although Eq.~\eqref{eq:learnable} makes our personalized subgraph selection learnable, it still cannot resolve the second challenge. This is because it only learns weight vectors $\{\alpha_{u,v}: e_{u,v}\in\mathcal{E}\}$ for observed edges, yet cannot generate weight variables for unseen ones.
As a result, it cannot be applied to infer missing edges.  

\subsubsection{Inductive Subgraph Selector.} To address the second challenge, we propose to make Eq.~\eqref{eq:learnable} inductive by computing the weight vector $\alpha_v$ with a deep neural network. Specifically, we estimate the contribution score $\alpha_{u,v}^{i,j}$ of edge $e_{u,v}$ \textit{w.r.t.} the subgraph $\mathcal{G}_{u,v}^{i,j}$ via a score function $g_{\bm{\theta}}:\mathcal{G}_{u,v}\xrightarrow{}\mathbb{R}$. Here $g_{\bm{\theta}}$ takes edge subgraphs as input and outputs their selection scores. Following this principle, we rewrite Eq.~\ref{eq:learnable} into an inductive version: 
\begin{equation}
\begin{split}
    \mathcal{G}_{u,v}=\sum_{1\leq{i,j} \leq K}\frac{\exp (g_\theta(\mathcal{G}_{u,v}^{i,j})/\tau)}{\sum_{i',j'=1}^K \exp(g_\theta(\mathcal{G}_{u,v}^{i',j'})/\tau)} * \mathcal{G}_{u,v}^{i,j}.
\end{split}
\label{eq:inductive}
\end{equation}
The above equation provides a principled solution to our personalized subgraph problem. On the one hand, it allows efficient subgraph selection for different edges based on the simple forward pass of a neural network. On the other hand, it enables the selection of the most informative subgraphs for unseen edges based on their subgraph characteristics. 
We now illustrate how to implement the score function $g_\theta$. Since $g_\theta$ takes $K^2$ subgraphs $\{\mathcal{G}_{u,v}^{i,j}: 1 \leq i,j\leq K\}$ of an edge $e_{u,v}$ as input, an intuitive strategy is applying GNNs to encode the $K^2$ subgraphs independently. However, this schema is time-consuming in training because it requires roughly $K^2$ GNNs forward passes to iterate over all edges once. 

\noindent\textbf{{Efficient $K^2$ subgraphs embedding}.} To avoid running GNNs forward pass repeatedly, we propose to get the representations of $K^2$ edge subgraphs simultaneously, by directly combining the hidden representations of end nodes from $K$ GNN layers. 
Specifically, given the $K$ hidden representations of node $u$ $\{\mathbf{h}_u^k : 1\leq k \leq K\}$ and $v$ $\{\mathbf{h}_v^k: 1\leq k \leq K\}$, we generate the representation of edge $e_{u,v}$ in terms of subgraph $(\mathcal{G}_u^i, \mathcal{G}_v^j)$ as $\mathbf{z}_{u,v}^{i,j}=\text{COM}(\mathbf{h}_u^i, \mathbf{h}_v^j)$. $\text{COM}(\cdot,\cdot)$ is a combination function, and the default setting is the element-wise multiplication. This approximation is reasonable in GNNs since $\mathbf{h}_v^i$ is obtained by aggregating messages from $v$'s neighbors within $i$ hops. \textbf{By doing this, we don't need to extract $k^2$ subgraphs and apply GNN on these graphs separately. Instead,  we can directly obtain $K^2$ subgraph embeddings upon the hidden representations of one GNN forward. So, the computational complexity of plugging in our personalized selector is close to standard GNNLPs (See Section~\ref{sec:efficiency} for efficiency analysis).} 
\label{efficient_trick}

After obtaining $K^2$ edge representations, we feed them into a MLP layer with ReLU activation function to predict their importance scores, i.e., $\alpha_{u,v}^{i,j}=g_{\bm{\theta}}(\mathbf{z}_{u,v}^{i,j})$. Through relaxing the hard selection operation, we can rewrite the mixed selection process in Eq.~\eqref{eq:inductive} in embedding space as:
\begin{equation}
\begin{split}
    \mathbf{z}_{u,v}=\sum_{1\leq {i,j} \leq K}\frac{\exp (g_\theta(\mathbf{z}_{u,v}^{i,j})/\tau)}{\sum_{i',j'=1}^K \exp(g_\theta(\mathbf{z}_{u,v}^{i',j'})/\tau)} \mathbf{z}_{u,v}^{i,j}.
\end{split}
\label{eq:emb}
\end{equation}
$\mathbf{z}_{u,v}\in\mathbb{R}^d$ is the final representation of edge $e_{u,v}$.
It is a mixed representation obtained by summing over the representations of various subgraph forms. Note that the mixed operation is only applied for the search phase. In the application phase, we output one subgraph for each edge via max selection (see Section 4).

\subsection{Model Training}
After generating the edge representation $\mathbf{z}_{u,v}$, we adopt an edge-wise loss function to estimate the reconstruction errors, expressed as:
\begin{equation}
\begin{split}
    \mathcal{L}=-\sum_{(v,u)\in\mathcal{E}} \frac{\exp({y}_{u,v})}{\sum_{u'\in\mathcal{V}}\exp({y}_{v,u'})},
\end{split}
\label{eq:loss}
\end{equation}
where ${y}_{u,v}=q_w(\mathbf{z}_{u,v})$ is the predicted score for edge $e_{u,v}$, and $q_w$ is another multilayer perceptron with ReLU activation. As the sum operation in the denominator of Eq.~\eqref{eq:loss} is computationally expensive, we adopt negative selection techniques~\cite{hamilton2017inductive} to accelerate the optimization in experiments. 

In the search phase, our goal is to jointly learn the subgraph selector $g_\mathbf{\bm{\theta}}$ and the model weights within the mixed selection, including GNN encoder $f_w$ and link predictor $q_w$. Following~\cite{zoph2018learning,liu2018darts}, we employ the validation set performance as a reward to optimize the subgraph selector, and train the GNN encoder and predictor by fitting the training set. Specifically, we optimize our model via the following bi-level optimization framework:  
\begin{equation}
\begin{split}
    \min_{g_{{\theta}}} \ 
    \mathcal{L}_{valid}(w^*, {\theta}) \ \ 
     \text{s.t.} \ \  w^*= \arg\min_{w}\mathcal{L}_{train}({\theta}^*, w).
\end{split}
\label{eq:bilevel}
\end{equation}
We use $w$ to wrap up the parameters of GNN encoder $f_w$ and link predictor $q_w$ for simplicity. $\mathcal{L}_{train}$ and $\mathcal{L}_{valid}$ denote the loss function in Eq.~\eqref{eq:loss} computed based on the training and validation sets, respectively. The upper-level objective $\mathcal{L}(w^*,{\theta})$ aims to find ${\theta}$ that minimizes the validation rewards given the optimal $w^*$, and the lower-level objective $\mathcal{L}({\theta}^*, w)$ targets to optimize $w$ by minimizing the training loss with ${\theta}$ fixed. 

It is worth noting that Eq.~\eqref{eq:bilevel} only exploits the cheap signals from observed edges, without accessing downstream labeled data for evaluation. Therefore, the validation set used to train the selector can be easily constructed. Since a closed-form solution cannot be computed, we optimize Eq.~\eqref{eq:bilevel} via alternating between the lower-level and the upper-level objectives as below. 

\subsubsection{Lower-level optimization.} With ${\theta}$ fixed, the lower-level optimization \textit{w.r.t.} $w$ follows the conventional gradient descent procedure, represented as:
\begin{equation}
\begin{split}
    w^\prime = w - \lambda \nabla_w\mathcal{L}_{train}(w, {\theta}^*),
\end{split}
\label{eq:lower}
\end{equation}
where $\lambda\in\mathcal{R}_{>0}$ is the learning rate. The converged solution is denoted as $w^*(\theta)$.

\subsubsection{Upper-level optimization.} With $w$ fixed, the upper-level optimization updates $\bm \theta$ according to the validation performance as:
\begin{equation}
\begin{split}
    \theta^\prime = \theta - \lambda \nabla_\theta\mathcal{L}_{valid}(w^*(\theta), \theta).
\end{split}
\label{eq:upper}
\end{equation}

However, evaluating the gradient \textit{w.r.t.} $\theta$ exactly is computationally prohibitive, since it requires solving for the optimal $w^*(\theta)$ whenever $\theta$ gets updated. To approximate the optimal solution $w^*(\theta)$, we propose to take one step of gradient descent update for $w$, without solving the lower-level optimization completely by training until convergence. The full derivation is delegated to Appendix~\ref{sec:C}. Here, we directly present the final result: 

\begin{equation}
    \begin{split}
        \nabla_\theta \mathcal{L}_{valid}(w^*(\theta), \theta)&\approx \nabla_\theta \mathcal{L}_{valid}(w^\prime, \theta) \\
        &- \lambda \frac{\nabla_{\theta}\mathcal{L}_{train}(w^+,\theta)-\nabla_{\theta}\mathcal{L}_{train}(w^-,\theta)}{2\epsilon},
    \end{split}
\label{weights_update2}
\end{equation}
where $w^\pm=w\pm\epsilon\nabla_{w^\prime}\mathcal{L}_{valid}(w^\prime(\theta), \theta)$, and $\epsilon$ is a small scalar for finite difference approximation. By alternating between the two update rules in Eq.~\eqref{eq:lower} and Eq.~\eqref{eq:upper}, we can learn an effective personalized subgraph selector that generalizes well for unseen edges. 
Although an optimizer with the theoretical guarantee of convergence for the bi-level optimization problem in Eq.~\eqref{eq:bilevel} remains an open challenge, alternating gradient descent algorithm has been widely adopted to solve similar objectives in Bayesian optimization~\cite{snoek2012practical}, automatic differentiation~\cite{zha2022towards}, and adversarial training~\cite{wang2019towards}. Algorithm~\eqref{alg:example} in Appendix depicts the optimization procedure of our model. It shows some level of empirical convergence as seen in Figure~\ref{train_losses} of Appendix.

\section{Application Phase}
After the search phase, we can apply the \textit{selected subgraphs} of different edges to various GNNLP models. In this section, we elaborate on two scenarios as examples. First, we illustrate how to train node2link-based models based on the selected edge subgraphs. Second, we show how to train subgraph2link-based methods given the sampled subgraphs. 

\subsection{Personalized node2link Based Models}
Typical examples under the node2link approach include GAE, GraphSAGE, LightGCN, and NGCF, to name a few. Given the subgraph $\mathcal{G}_{u,v}=(\mathcal{G}_v^i, \mathcal{G}_u^j)$ of edge $e_{u,v}$, the models continue to learn node representations of end node subgraphs using GNNs, and then combine the representations of end nodes as the edge embedding towards prediction. In traditional settings, $i=j=k$, the edge representation can be easily generated by concatenating embeddings of end nodes in the last GNN layer, i.e., $\mathbf{z}_{u,v}=[\mathbf{h}_u^k, \mathbf{h}_v^k]$. When using personalized subgraph selection, where $i$ and $j$ could be different, we compute the personalized edge embedding via $\mathbf{z}_{u,v}=[\mathbf{h}_u^i, \mathbf{h}_v^j]$. 
\begin{figure}[H]
\vspace{-5pt}
  \centering
  \includegraphics[scale=0.5]{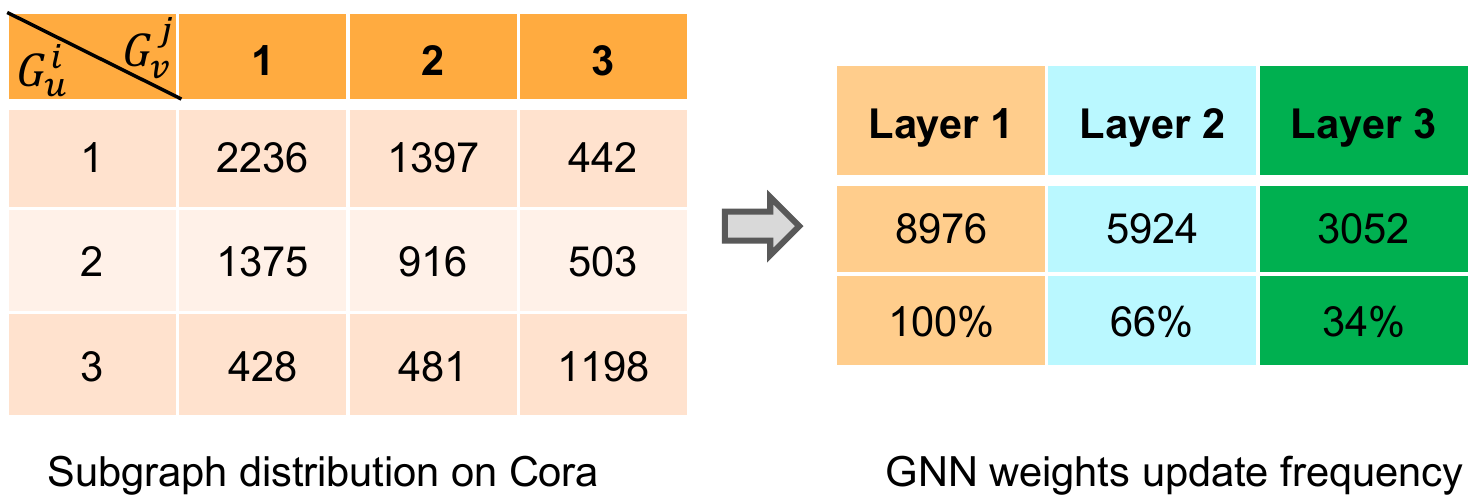}\\
  \caption{An example illustration of the subgraph imbalance issue. The frequency of updating three GNN layers differs due to the imbalanced edge distribution across layers.}
  \label{fig:imbalance}
  \vspace{-5pt}
\end{figure}
\subsubsection{Subgraph Imbalance Issue.} However, directly training existing node2link-based models over the personalized edge subgraphs may suffer from the subgraph imbalance issue shown in Figure~\ref{fig:imbalance}. We can observe that the three GNN layers will be updated inconsistently in mini-batch training, since the number of edges being encoded across three layers is different. To eliminate this issue, inspired by the success of the pre-training models in texts~\cite{devlin2018bert} and images~\cite{dosovitskiy2020image}, we adopt the pre-train\&finetune fashion to train node2link methods in personalized subgraph setting. By initializing the GNN encoder of the application model with the pre-trained one in the search phase, the application model can be well-tuned with limited training samples (a.k.a. limited training edges in the third layer). Note that the subgraph imbalance issue is different from class imbalance problem~\cite{longadge2013class} in standard machine learning, since the training subgraphs in deeper layers are dependent on previous layers, which makes up-sampling or down-sampling techniques not applicable.

\subsection{Personalized subgraph2link Based Models}
Different from node2link-based methods, subgraph2link approaches naturally take the subgraphs of anchor edges as input, since they treat edge embedding as a graph-level representation learning task. Therefore, the personalized edge subgraphs generated by our model can be directly fed to them as input without additional effort.

\section{Experiments}
We try to answer five research questions through experiments. 
\textbf{Q1:} Is considering personalized edge subgraphs beneficial for GNNLP models when evaluated on different applications? \textbf{Q2:} How effective is the proposed personalized subgraph selector in identifying edge subgraphs across various datasets? 
\textbf{Q3:} How will our personalized subgraph selector react to the changes in different optimization strategies? 
\textbf{Q4:} What are the impacts of hyperparameters on \model, such as the maximum hop number $K$ and the embedding dimension $D$ of the score function? \textbf{Q5:} What is the running complexity of our personalized subgraph selector compared with standard GNNLPs?

\subsection{Datasets and Experiment Settings}
\textbf{Datasets.} For a comprehensive comparison, we use nine datasets of diverse nature with both homogeneous and heterogeneous graphs. For homogeneous graphs, we consider six popular datasets including three benchmark Planteoid datasets (\textbf{Cora}, \textbf{CiteSeer}, and \textbf{PubMed}~\cite{sen2008collective}) and three large-scale benchmark datasets, \textbf{ogbl-ddi}, \textbf{ogbl-collab}, and \textbf{ogbl-ppa} from Open Graph Benchmark (OGB)~\cite{hu2021ogb}. We summarize their statistics in Table~\ref{dataset_stat_homo} of Appendix. For heterogeneous graphs, we include three benchmark datasets including \textbf{Gowalla}~\cite{liang2016modeling}, \textbf{Yelp2018}~\cite{wang2019neural}, and \textbf{Amazon-book}~\cite{he2016ups}. We summarize their statistics in Table~\ref{dataset_stat_hete} of Appendix. 

\noindent\textbf{Learning protocols.} We aim to provide a rigorous and fair comparison between different models across various graph domains by following the standard dataset splits and training procedure. For homogeneous graphs, we follow~\cite{kipf2016variational} to randomly split three graphs in Planetoid datasets into three sets, i.e., the training set (85\%), the validation set (5\%), and the test set (10\%), and measure model performance based on AUC and Average Precision (AP) scores. For OGB datasets (ogbl-ddi, ogbl-collab, and ogbl-ppa), we follow~\cite{hu2020open} to split the datasets into three sets according to the split ratio summarized in Table~\ref{dataset_stat_homo}, and evaluate the performance using Hit rate (Hit@$N$), where $N$ is the number of nodes recalled. For heterogeneous graphs, we follow~\cite{he2020lightgcn} to generate the training, validation, and testing sets with split ratios in Table~\ref{dataset_stat_hete}. Since it is too time-consuming to rank all items for every user during evaluation, we follow the common strategy~\cite{he2017neural} that randomly samples 100 items that are not interacted with by the user, ranking the test item among the sampled items. The performance of the ranked list is judged by two widely-used evaluation metrics: hit@$N$ and ndcg@$N$.

\begin{table*}[htbp]
\caption{Link prediction results on Planetoid data. "+"/"-" in the bracket indicates relative improvement with baselines. }
  \vspace{-8pt}
 \begin{small}
\setlength{\tabcolsep}{1.5pt}
{
\begin{tabular}{l cc cc cc cc}
\toprule
  &\multicolumn{2}{c}{Cora}
 &\multicolumn{2}{c}{CiteSeer}
 &\multicolumn{2}{c}{PubMed}\\
\cmidrule(r){2-3} \cmidrule(r){4-5} \cmidrule(r){6-7}
 &AUC &AP &AUC &AP
&AUC &AP\\

\cmidrule(r){1-3} \cmidrule(r){4-5} \cmidrule(r){6-7}
GAE &$91.08 \pm 0.01 $ &$92.03 \pm 0.03 $ &$89.52\pm 0.04 $ &$89.95\pm 0.05 $ &$96.40 \pm 0.01 $&$96.50 \pm 0.02$\\
GAE-RS &$89.38\pm0.23 (-1.8\%)$ &$91.66\pm0.45 (-0.4\%)$ &$89.81\pm0.29 (0.3\%)$ &$91.39\pm0.22 (1.6\%)$ &$95.34\pm0.23 (-1.1\%)$ &$96.13\pm0.18 (-0.3\%)$\\
GAE-PS2 &$92.25\pm0.71 (+1.4\%)$ &$93.60\pm0.25 (+1.7\%)$ &$92.16\pm0.19 (+3.0\%)$ &$93.07\pm0.06 (+3.6\%)$ &$98.27\pm0.10 (+2.0\%)$ &$98.10\pm0.16 (+1.7\%)$\\
\midrule
GraphSage &$86.36 \pm 1.06$ &$88.22 \pm 0.87$ &$85.24\pm 2.56$ &$86.60\pm 2.54$ &$87.61\pm0.87$&$89.41\pm0.82$\\
GraphSage-RS &$89.27\pm0.70 (+3.3\%)$ &$89.67\pm0.58 (+1.6\%)$ &$88.30\pm1.08 (+3.5\%)$ &$89.23\pm0.91 (+3.0\%)$ &$88.72\pm0.77 (+1.3\%)$ &$90.58\pm0.75 (+1.3\%)$\\
GraphSage-PS2 &$93.86\pm0.01 (8.7\%)$ &$93.35\pm0.25 (5.8\%)$ &$93.46\pm1.20 (9.6\%)$ &$93.62\pm1.10 (8.1\%)$ &$93.47\pm0.22 (6.7\%)$ &$93.52\pm0.25 (4.6\%)$\\
\midrule
SEAL &$90.82\pm1.97$&${92.18\pm0.82}$&$88.49\pm1.22$&$90.64\pm1.46$&${97.57\pm0.05}$&${97.20\pm0.03}$\\
SEAL-RS &$88.55\pm0.88 (-2.5\%)$ &$90.37\pm0.36 (-1.9\%)$ &$86.70\pm1.34 (-2.0\%)$ &$88.61\pm1.62 (-2.2\%)$ &$94.39\pm0.35 (-3.2\%)$ &$94.08\pm0.27 (-3.2\%)$\\
SEAL-PS2 &$92.31\pm0.31 (+1.6\%)$ &$93.53\pm0.22 (+1.4\%)$&$90.29\pm0.36 (+2.0\%)$&$92.42\pm0.20 (+1.9\%)$&$97.60\pm0.10 (+0.0\%)$&$97.44\pm0.05 (+0.2\%)$\\
\bottomrule
\end{tabular}}
 \end{small}
\label{table_lp}
\vspace{-4pt}
\end{table*}

\begin{table*}[htbp]
\caption{Link prediction results on recommendation datasets.}
  \vspace{-4pt}
 \begin{small}
\setlength{\tabcolsep}{1.5pt}
{
\begin{tabular}{l cc cc cc cc}
\toprule
  &\multicolumn{2}{c}{Gowalla}
 &\multicolumn{2}{c}{Yelp}
 &\multicolumn{2}{c}{Amazon-book}\\
\cmidrule(r){2-3} \cmidrule(r){4-5} \cmidrule(r){6-7}
 &Hit@10 &ndgc@50 &Hit@10 &ndgc@50
&Hit@10 &ndgc@50\\

\cmidrule(r){1-3} \cmidrule(r){4-5} \cmidrule(r){6-7}
NGCF &$85.16\pm 0.58$ &$64.20 \pm 0.34$ &$80.17\pm 0.61$ &$55.12\pm0.44$ &$70.15\pm0.59$ &$50.63\pm0.64$\\
NGCF-RS &$83.55\pm0.47(-1.8\%)$ &$62.74\pm0.40(-2.2\%)$ &$78.07\pm0.53(-2.6\%)$ &$52.90\pm0.32 (-4.0\%)$ &$67.82\pm0.65 (-3.3\%)$ &$48.02\pm0.81 (-5.1\%)$\\
NGCF-PS2 &$88.21\pm0.24 (+3.6\%)$ &$65.59\pm0.60 (+1.4\%)$ &$87.72\pm0.35 (+9.4\%)$ &$56.61\pm0.27 (+2.7\%)$ &$78.66\pm0.30 (+12.1\%)$ &$51.43\pm0.41 (+1.6\%)$\\
\midrule
LightGCN &$86.96\pm0.46$ &$68.11\pm0.18$ &$82.44\pm0.37$ &$58.94\pm0.28$ &$74.32\pm0.36$ &$54.79\pm0.24$\\
LightGCN-RS &$85.85\pm0.39 (-1.3\%)$ &$66.69\pm0.20 (-2.1\%)$ &$81.72\pm0.57 (-0.9\%)$ &$57.88\pm0.35 (-1.8\%)$ &$73.75\pm0.38 (-0.8\%)$ &$53.51\pm0.18 (-2.3\%)$\\
LightGCN-PS2 &$89.79\pm0.25 (+3.3\%)$ &$69.75\pm0.20 (+2.4\%)$ &$88.44\pm0.32 (+7.3\%)$ &$59.99\pm0.19 (+1.8\%)$ &$81.26\pm0.26 (+9.3\%)$ &$59.29\pm0.33 (+8.2\%)$\\
\bottomrule
\end{tabular}}
 \end{small}
\label{table_lp_rec}
\vspace{-4pt}
\end{table*}

\noindent\textbf{Baselines.} To demonstrate the effectiveness, we compare our model with state-of-the-art link prediction methods of two domains. For homogeneous graphs, we include two popular node2link based methods (\textbf{GAE}~\cite{kipf2016variational} and \textbf{GraphSage}~\cite{hamilton2017inductive}) and one subgraph2link based method (\textbf{SEAL}~\cite{zhang2018link}). For heterogeneous graphs, we consider two recently proposed benchmark methods (\textbf{NGCF}~\cite{wang2019neural} and \textbf{LightGCN}~\cite{he2020lightgcn}). Besides, we include one variant of our model based on the random search, named "RS". For all baseline methods, we use their open-source implementations with the best configurations on datasets that are tested in original papers. For datasets not originally tested, we tune their hyperparameters according to the range suggested in original papers. 

\noindent\textbf{Implementation details.} Our model is built upon the Pytorch platform. We train our model for 100 epochs with Adam optimizer and early stopping with patience of 20 epochs. Following common practice in~\cite{he2020lightgcn,kipf2016variational,hu2020open}, we employ a three-layer GNN encoder with dimension 32, 256, and 64 for the Planetoid, OGB, and heterogeneous datasets, respectively. When applying our personalized subgraph selector PS2 to node2link-based baselines (GAE, GraphSage, NGCF, and LightGCN), we use the same GNN architectures as the vanilla counterparts in the search phase. For subgraph2link-based baseline (SEAL), we employ GCN~\cite{kipf2016semi} as the backbone in the search phase, since it is memory and time expensive to generate subgraph embeddings by pooling over the whole subgraph as SEAL does.  Our model has two hyper-parameters, i.e., the maximum hop number $K$ and the hidden dimension $D$ of score function $g_\theta$. We set $K=3$ by default and search $D$ within the set $\{64, 128, 256, 512, 1024\}$. The best options for three Planetoid and other datasets are 256 and 512, respectively. We provide more details in {Appendix~\ref{sec:B}}.

\subsection{Comparison with the Baselines}
To answer the question \textbf{Q1}, we compare the performance of the proposed personalized subgraph selector with state-of-the-art baselines across homogeneous and heterogeneous domains. Table~\ref{table_lp}, Table~\ref{table_lp_ogb} and Table~\ref{table_lp_rec} report the results over Planetoid, OGB, and three recommendation datasets, respectively. From the tables, we have the following \textbf{Obs}ervations. 

\textbf{Obs. 1. Through edge subgraph personalization, PS2 boosts the performance of link prediction across different domains.} By comparing classical GNNLP methods (GAE, GraphSage, SEAL, NGCF, and LightGCN) with our personalized subgraph selector (GAE-PS2, GraphSage-PS2, SEAL-PS2, NGCF-PS2, and LightGCN-PS2), our model consistently outperforms the vanilla counterparts on both homogeneous and heterogeneous graphs (in Table~\ref{table_lp} and Table~\ref{table_lp_rec}). Specifically, on homogeneous graphs (Table~\ref{table_lp}), GAE-PS2, GraphSage-PS2, and SEAL-PS2 achieve better results than GAE, GraphSage, and SEAL across two evaluation metrics. Our model has different impacts concerning various backbones. For example, GraphSage-PS2 significantly outperforms GraphSage with up to 9.6\% improvements. In heterogeneous scenarios, NGCF-PS2 and LightGCN-PS2 generally perform better than NGCF and LightGCN on three datasets. In particular, the performance gap between our model and two baselines increases on top-$10$ based metrics. This result verifies the effectiveness of our model in accurately recalling related items in the top-ranking list. 

\textbf{Obs. 2. Across various datasets, the proposed personalized subgraph selector consistently outperforms random search-based variants.}
For different datasets and scenarios (in Table~\ref{table_lp} and Table~\ref{table_lp_rec}), our model consistently outperforms the random search based variants with a large margin. Specifically, random search-based variants are not robust across various datasets. For example, GraphSage-RS could generally achieve better or comparable results with their counterparts on Cora, CiteSeer, and PubMed datasets. But it loses to their counterparts on recommendation datasets in all cases (see Table~\ref{table_lp_rec}). This comparison validates our motivation to design an automated subgraph selector in a data-driven fashion.

\begin{table}[htp]
\centering
\caption{Link prediction performance on OGB datasets.}
  \vspace{-13pt}
 \begin{small}
\setlength{\tabcolsep}{1pt}
{
\begin{tabular}{l c c c}
\toprule
  &{ogbl-ddi}
 &{ogbl-collab}
 &{c}{ogbl-ppa}\\
   &{Hit@20}
 &{Hit@50}
 &{Hit@100}\\
\midrule
GAE &$37.07 \pm 5.07 $ &$44.75\pm 1.07 $ &$18.67 \pm 1.32$\\
GAE-RS &$30.72\pm6.86$ &$34.51\pm0.68$ &$12.65\pm1.22$\\
GAE-PS2 &$\bf{49.53\pm5.99}$ &$\bf{50.26\pm0.32}$ &$\bf{20.50\pm0.79}$\\
\midrule
GraphSage &$53.90 \pm 4.74$ &$54.63\pm 1.12$ &$16.55\pm2.40$\\
GraphSage-RS &$27.17\pm5.74$ &$37.54\pm0.37$ &$9.89\pm3.46$\\
GraphSage-PS2 &$\bf{56.90 \pm 5.32}$ &$\bf{55.71 \pm 0.93}$ &$\bf{17.88\pm1.33}$\\
\midrule
SEAL &$30.56\pm3.86$&$63.64\pm0.71$&$48.80\pm3.16$\\
SEAL-RS &$24.58\pm4.65$ &$43.56\pm1.30$ &$35.68\pm5.21$\\
SEAL-PS2 &$\bf{32.77\pm2.50}$ &$\bf{64.83\pm0.54}$ &$\bf{50.25\pm2.33}$\\
\bottomrule
\end{tabular}}
 \end{small}
\label{table_lp_ogb}
\vspace{-5pt}
\end{table}

\textbf{Obs. 3. The proposed PS2 scales up well on large-scale datasets.} On three challenging OGB datasets, our model PS2 can continuously boost the performance of vanilla GNNLP methods, as shown in Table~\ref{table_lp_ogb}. Specifically, GAE-PS2 improves 60.5\%, 14.5\%, and 9.8\% over GAE on ogbl-ddi, ogbl-collab, and ogbl-ppa datasets, respectively. In contrast, the random search-based variants lose to their counterparts on these three datasets. This observation further demonstrates the effectiveness of considering learnable subgraph selection on large graphs. 

\begin{figure}[htp]
  \vspace{-6pt}
  \includegraphics[width=8 cm, height=4 cm]{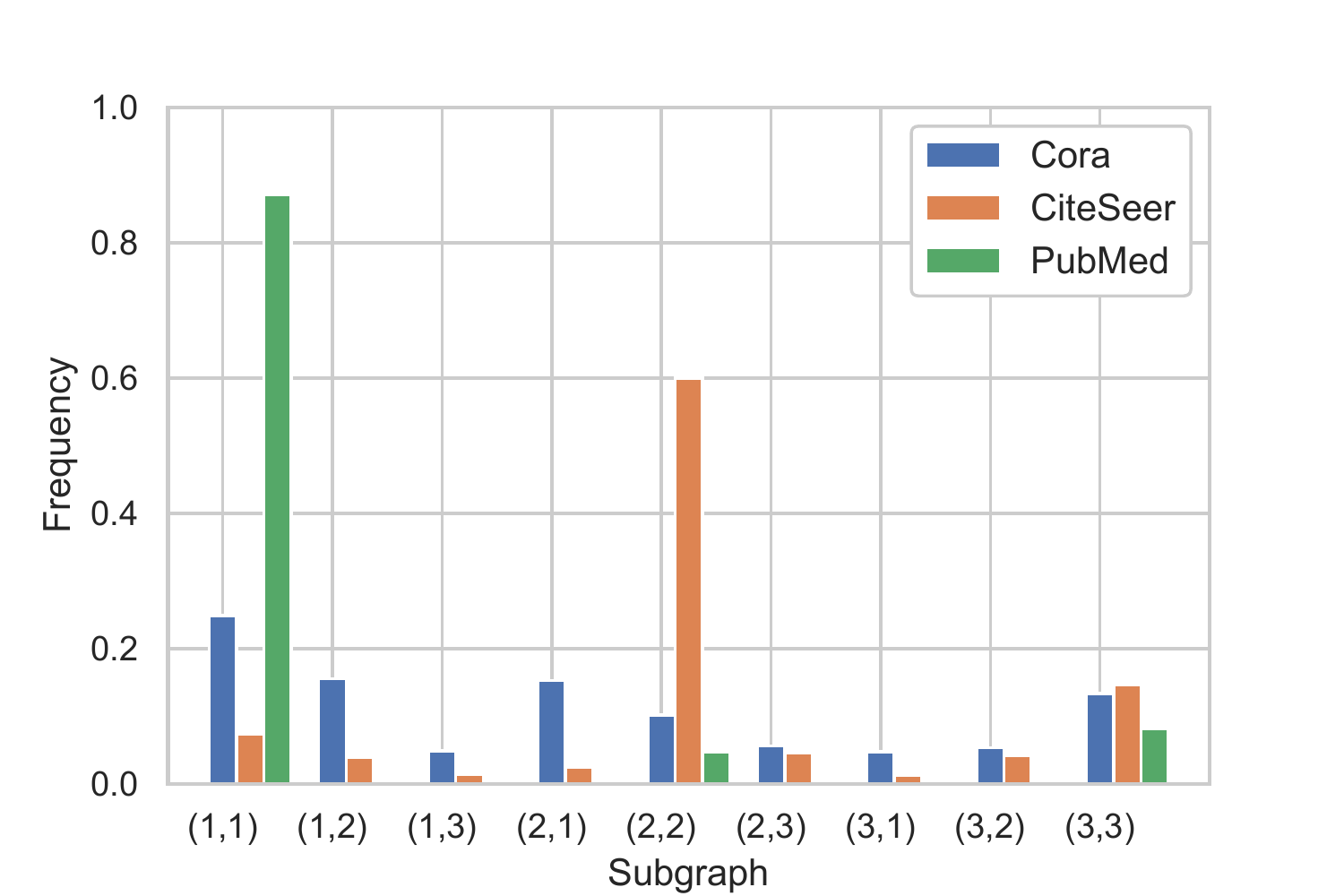}
  \caption{Subgraph distribution of GAE-PS2 on the Plantetoid.
  }
  \label{fig:subgraph}
  \vspace{-28pt}
\end{figure}

\subsection{Subgraph Distribution Analysis}
We visualize the learned subgraph distributions of GAE-PS2 on all datasets in Figure~\ref{fig:subgraph} and Figure~\ref{fig:subgraph_appenx} in Appendix to study (\textbf{Q2}). By comparing the distributions across different benchmarks, we have the following observation.

\textbf{Obs. 4. By learning from the data, PS2 can effectively learn different subgraph distributions for various datasets, and even skip some suboptimal subgraphs.} Our model PS2 can identify different subgraphs for different edges, and allow different datasets to have their own subgraph distributions (see Figure~\ref{fig:subgraph}). Specifically, the subgraph distribution on OGB datasets is more sparse than the other two types of datasets, while recommendation datasets generally tend to have smoother distribution. One promising property of PS2 is that it can skip some subgraphs if they are not optimal for any edges. For example, no edges are assigned to the subgraph $(2,2)$ on ogbl-ddi and ogbl-ppa datasets. 

\begin{figure}[ht]
  \centering
  \includegraphics[width=7cm,height=3cm]{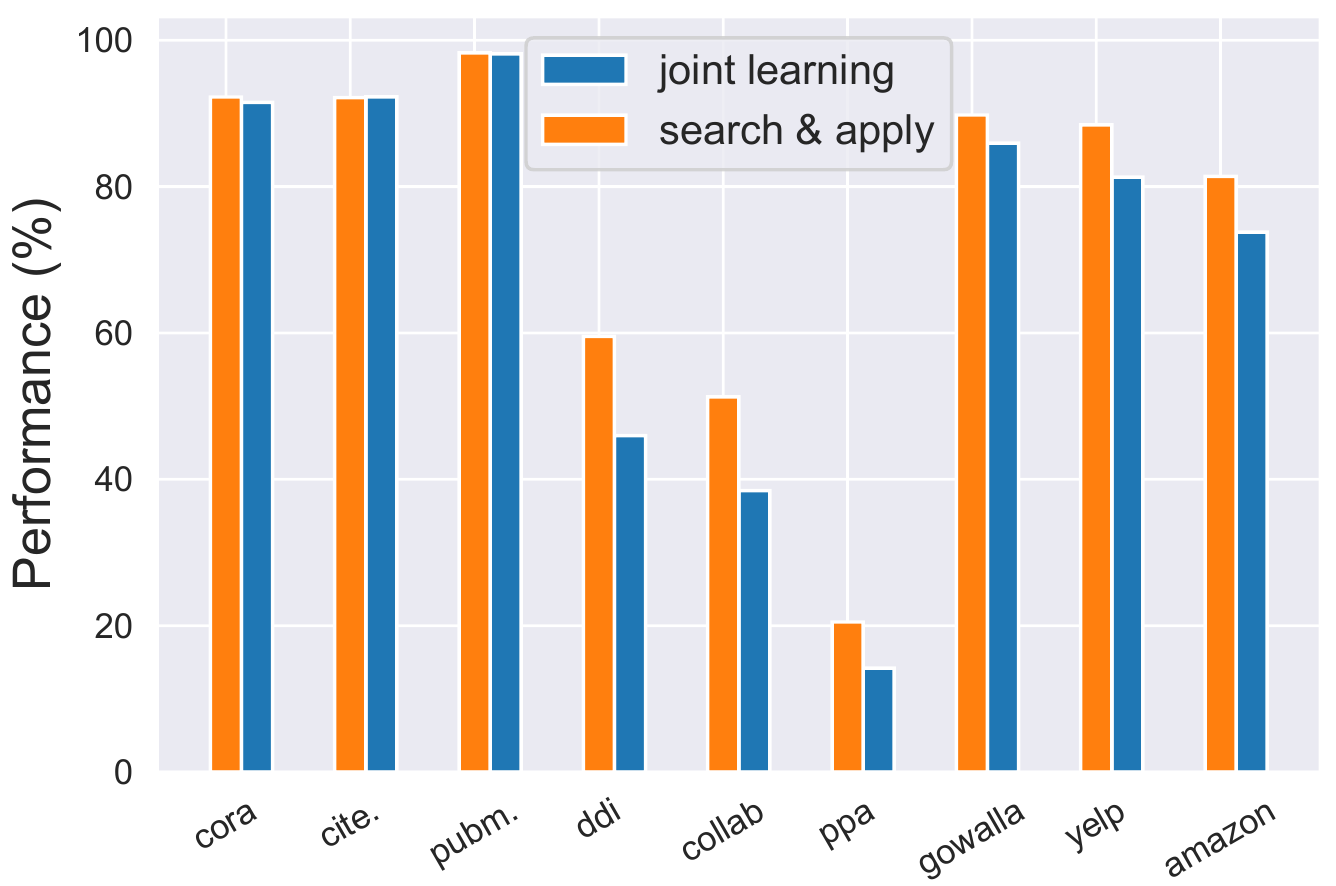}
  \vspace{-8pt}
  \caption{The performance of GAE-PS2 under different optimization paradigms. The evaluation metrics for (Cora, CiteSeer, and Pubmed) and other datasets are AUC and Hit ratio. 
  }
  \label{fig_op_ps2}
  \vspace{-8pt}
\end{figure}
\subsection{Optimization Analysis}
To examine the influence of optimization strategies on~\model~ (\textbf{Q3}), we compare the default \textit{search \& apply} paradigm with the \textit{joint learning} schema on node2link based models. Here, jointly learning means we directly train PS2 with a downstream inference model, i.e., GAE, end-to-end. In this setting, the PS2 training still uses the mixed selection, while the GAE optimization exploits the most likely subgraph via maximum discrete selection. Figure~\ref{fig_op_ps2} shows the results of two settings on GAE-PS2 over all datasets.  We can observe that although \textit{joint learning} strategy achieves comparable results with \textit{search \& apply} schema on Cora, CiteSeer, and PubMed datasets, the later schema performs better on the other six large-scale datasets. The possible reason is that joint learning is hard to optimize since the personalized selector and the downstream model are entangled. This comparison validates our choice to adopt the search \& apply fashion similar to the AutoML~\cite{liu2018darts} domain.  

Besides, we also explore the effectiveness of finetune strategy to avoid the subgraph imbalance issue when applying PS2 for node2link-based methods. Table~\ref{table_lp_self} in Appendix reports the results on Planetoid datasets. Similar observations could be made on other datasets. From Table~\ref{table_lp_self}, we observe that finetune strategy outperforms training from scratch on GAE and GraphSage backbones.

\begin{figure}[t]
  \centering
  \begin{subfigure}[b]{0.23\textwidth}
    \centering
    \includegraphics[width=0.8\textwidth]{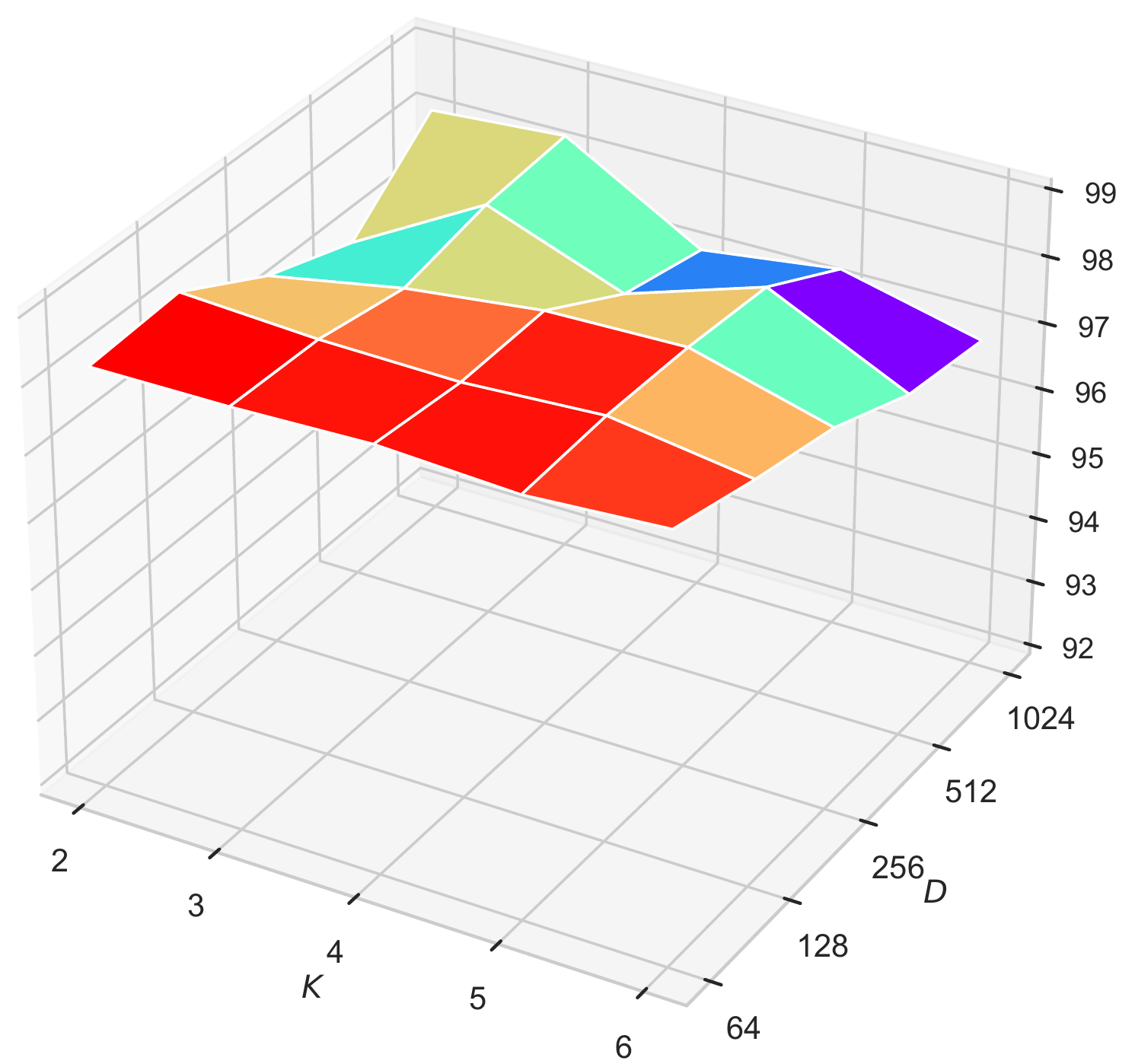}
    \vspace{-8pt}
  \end{subfigure}%
  \begin{subfigure}[b]{0.23\textwidth}
    \centering
    \includegraphics[width=0.8\textwidth]{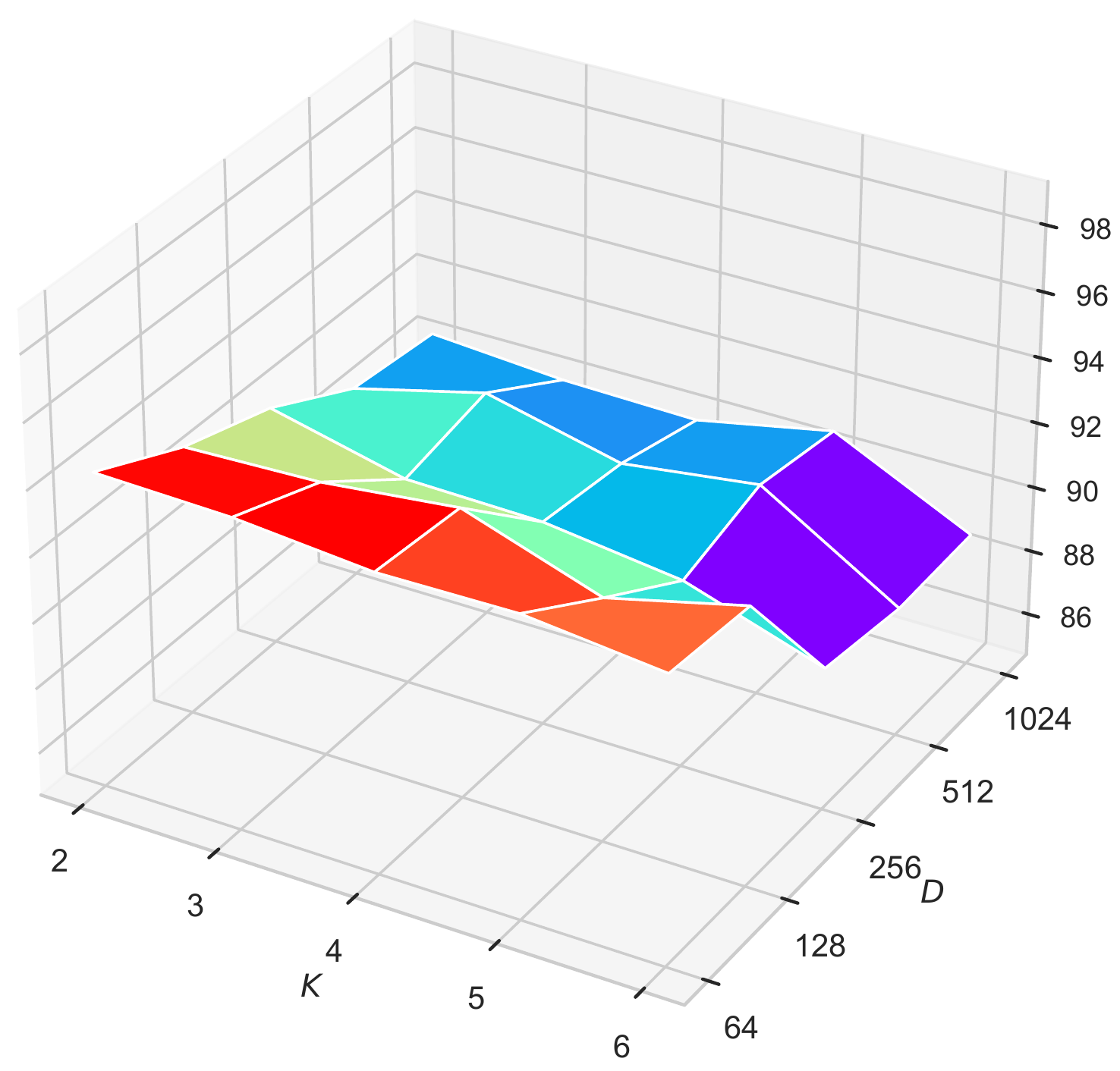}
    \vspace{-8pt}
  \end{subfigure}%
  \caption{Hyperparameter study of PS2 on BlogCatalog with different base models: GAE (left) and GraphSage (right).}
  \label{fig_hyper}
   \vspace{-12pt}
\end{figure}

\subsection{Hyperparameter Analysis}
To answer \textbf{Q4}, we study the impact of $K$ and hidden dimension $D$ of the score function on PubMed. Figure~\ref{fig_hyper} shows the results on GAE-PS2 and GraphSage-PS2. From the two subfigures, we can observe that our model performs relatively stable over a wide range of combinations of $K$ and $D$. Specifically, the best results in two cases are achieved when $K$ and $D$ are around 3 and 256, respectively. Similar observations are obtained on other datasets. In experiments, we fix $K=3$ and set $D=256$ and $D=512$ for Planetoid and other datasets (OGB and recommendation), respectively. 

\subsection{Efficiency Analysis}
\label{sec:efficiency}
To study \textbf{Q5}, we analyze the training costs of two representatives GNNLPs (GAE and SEAL) after plugging in our personalized selector. For SEAL, we exclude the sampling cost since it is far more than its forward pass running costs. From Table~\ref{table_efficiency} of Appendix, we observe that the additional costs to activate specific subgraph using our personalized selector is marginal, i.e., usually less than 20\% running consumption. This is because our selector is simple MLPs, and we can directly generate subgraph embeddings based on GNN output, thanks to the embedding approximation trick in Section~\ref{efficient_trick}. 



\section{Related Work}
In this paper, we mainly focus on graph neural networks (GNNs) based link prediction (GNNLP) techniques. For methods beyond GNN, please refer to~\cite{kumar2020link,zhou2021progresses} for a comprehensive review. For illustration purposes, the existing methods can be mainly divided into two categories:
node2link~\cite{kipf2016variational,hamilton2017inductive,tan2023} and subgraph2link~\cite{zhang2018link,zhang2021labeling,pan2021neural}.

\noindent\textbf{node2link} is the classical approach to perform link prediction based on GNNs. Given a query edge, it works by first generating representations for two end entities based on their local subgraphs via the GNNs encoder, and then combining the two representations to estimate the edge existence probability. Some efforts have been made to predict missing edges for homogeneous graphs~\cite{kipf2016variational,hamilton2017inductive,pan2018adversarially,ai2022structure}, while several methods propose to tackle link prediction on heterogeneous graphs, such as recommendation systems~\cite{ying2018graph,wang2019neural,he2020lightgcn,wu2020graph,zhou2021temporal} and knowledge graph completion~\cite{arora2020survey}. 

\noindent\textbf{subgraph2link} is a recently proposed new link prediction paradigm. The key idea is to represent each edge with a subgraph around it, and then apply GNNs to learn representation for the whole subgraph. The pioneering work of~\cite{zhang2018link} adopts node labeling to first create structure-aware features for nodes in the subgraph, and then pool over the node representations obtained by GNNs to get the final edge representation. A follow-up work~\cite{zhang2021labeling} analyzes the impacts of different labeling techniques. Another recent work~\cite{pan2021neural} proposes to replace the pooling operation with a more advanced yet complicated random-walk-based pooling strategy.    

\section{Conclusion}
In this paper, we explore a new perspective to train link prediction models by considering edge personalization in terms of neighborhood subgraphs. Specifically, we propose an effective personalized subgraph selector (PS2) as a plug-and-play framework for the graph neural network based link prediction (GNNLP) community. PS2 can automatically and inductively identify optimal subgraph orders for different edges when performing GNNLP. Extensive experiments on multiple datasets with various domains and scales demonstrate the superiority of PS2 against diverse GNNLP backbones. In the future, we will extend PS2 to perform subgraph order selection and critical neighbor sampling within the selected subgraph jointly. 

\begin{acks}
We thank the anomalous reviewers for the feedback. The work is, in part, supported by NSF (IIS-1849085, IIS-1750074, IIS-2006844). The views and conclusions in this paper are those of the authors and should not be interpreted as representing any funding agencies.
\end{acks}

\balance
\bibliographystyle{unsrt}
\bibliography{main}

\clearpage
\newpage
\appendix

\begin{figure*}[htp]
  \centering
  \begin{subfigure}[b]{0.33\textwidth}
    \centering
    \includegraphics[width=0.99\textwidth]{fig_dist_planetoid.pdf}
    \vspace{-8pt}
  \end{subfigure}%
  \begin{subfigure}[b]{0.33\textwidth}
    \centering
    \includegraphics[width=0.99\textwidth]{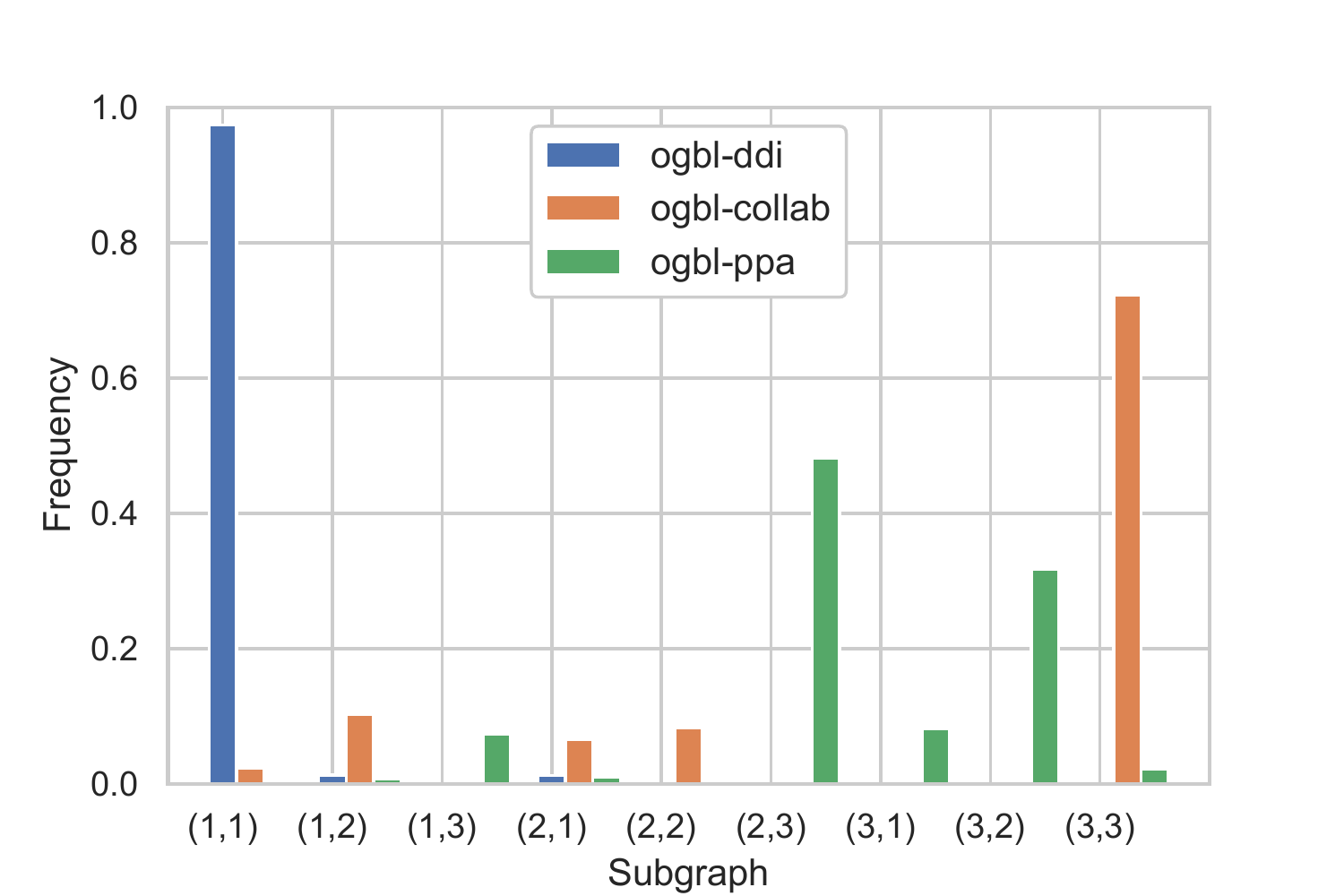}
    \vspace{-8pt}
  \end{subfigure}%
  \begin{subfigure}[b]{0.33\textwidth}
    \centering
    \includegraphics[width=0.99\textwidth]{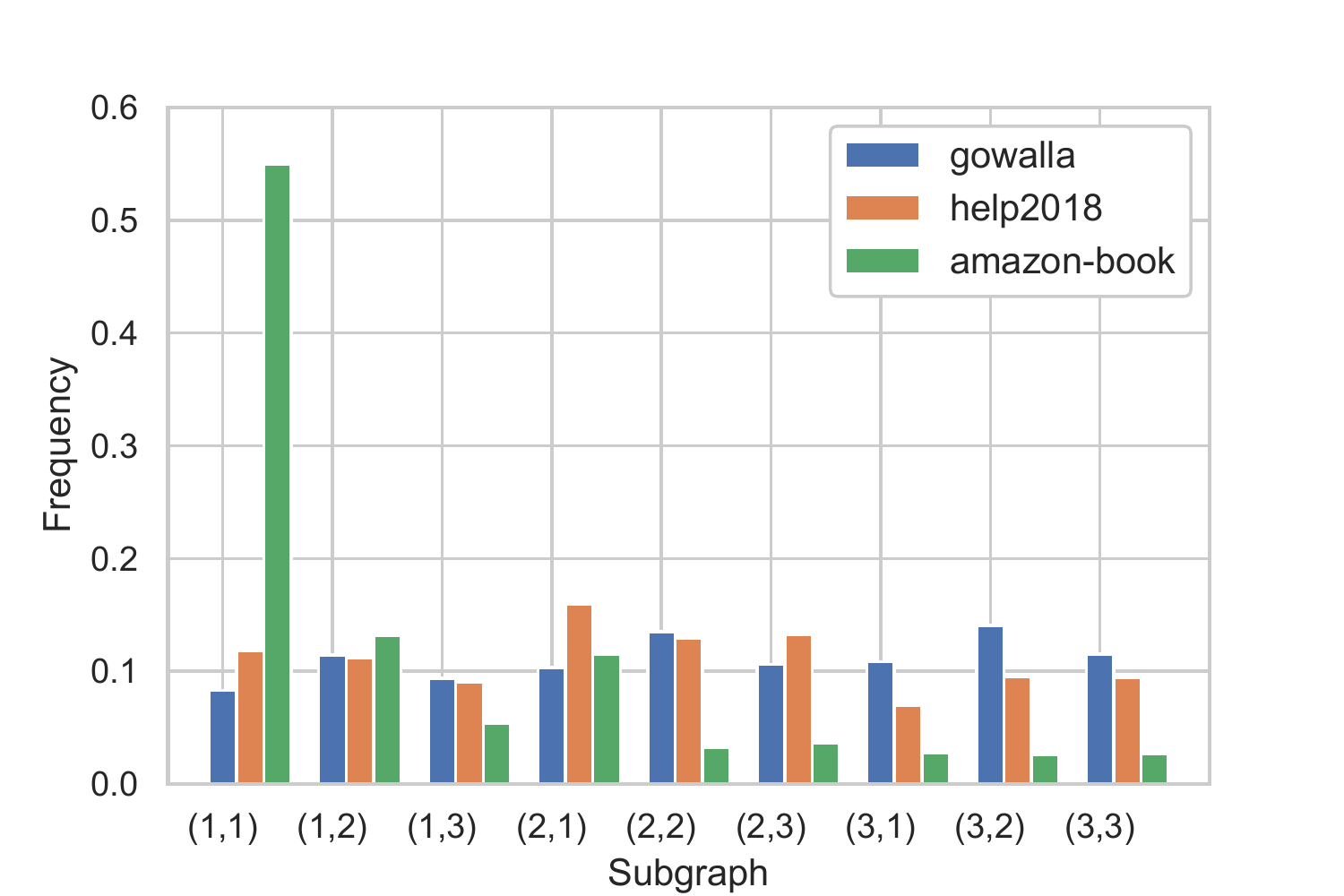}
    \vspace{-8pt}
  \end{subfigure}%
  \caption{Subgraph distribution of GAE-PS2 on the Planetoid (left), OGB (middle), and Recommendation (right) datasets.}
  \label{fig:subgraph_appenx}
\end{figure*}

\section{Dataset Details}
\label{sec:A}
In this section, we introduce the details of applied datasets as below.
\begin{itemize}
    \item \textbf{Cora, CiteSeer}, and \textbf{PubMed}: They are the most popular benchmark citation networks used in the graph domain. Nodes correspond to documents and edges correspond to citations. Each node has a bag-of-words feature vector according to the paper abstract. Labels are defined as academic topics.
    \item\textbf{ogbl-ddi}: This is a drug-drug interaction network. Each node represents an FDA-approved or experimental drug. Edges represent interactions between drugs. Node features are not available, in experiments, following~\cite{hu2021ogb}, we randomly initialize a 256-dimensional embedding vector for each node. 
    \item\textbf{ogbl-collab}: This is a challenging author collaboration network from KDD Cup 2021. Each node is an author and edges indicate the collaboration between authors. All nodes come with 128-dimensional features, obtained by averaging the word embeddings of papers that are published by the authors.
    \item\textbf{ogbl-ppa}: This is a protein-protein association network. Nodes represent proteins from 58 different species, and edges indicate biologically meaningful associations between proteins. In experiments, we use the 58-dimensional one-hot vectors as node features.  
\end{itemize}
In addition to the aforementioned six homogeneous graphs, we also consider three popular recommendation datasets.
\begin{itemize}
    \item \textbf{Gowalla}: This is the check-in dataset obtained from Gowalla, where users share their locations by checking-in. To ensure the qualify of the dataset, following~\cite{wang2019neural}, we use the 10-core setting, i.e., retaining users and items with at least ten interactions. 
    \item\textbf{Yelp2018}: This dataset is adopted from the 2018 edition of the Yelp challenge. It describes the relationships between customers and items like restaurants and bars. We use the same 10-core setting in order to ensure data quality.  
    \item\textbf{Amazon-book\footnote{\url{https://jmcauley.ucsd.edu/data/amazon/}}}: This is one of the widely used datasets for product recommendation. Similarly, we use the 10-core setting to ensure that each user and item have at least ten interactions. 
\end{itemize}

We split all datasets above into the training/validation/testing sets according to common practice~\cite{kipf2016variational,hu2021ogb,wang2019neural} and the specific splitting ratios are summarized in Table~\ref{dataset_stat_homo} and~\ref{dataset_stat_hete}.

\begin{table}[t]
\small
  \caption{Statistics of homogeneous graph datasets.}
    \vspace{-8pt}
  \begin{tabular}{c| c |c |c | c c}
   \toprule
     Data&\# Nodes &\# Edges &\# Features &Split ratio\\
     \hline
     Cora &$2,708$ &$5,429$ &$1,433$ &$85/5/15$ \\
     CiteSeer &$3,312$ &$4,660$ &$3,703$ &$85/5/15$ \\
     PubMed &$19,717$ &$44,338$ &$500$ &$85/5/15$ \\
     ogbl-ddi &$4,267$ &$1,334,889$ &- &$80/10/10$ \\
     ogbl-collab &$235,868$ &$1,285,465$ &$128$ &$92/4/4$ \\
     ogbl-ppa &$576,289$ &$30,326,273$ &$58$ &$70/20/10$ \\
 \bottomrule
\end{tabular}
\label{dataset_stat_homo}
\vspace{-8pt}
\end{table}

\begin{table}[t]
\small
  \caption{Statistics of heterogeneous graph datasets.}
    \vspace{-8pt}
  \begin{tabular}{c| c |c |c | c}
   \toprule
     Data&\# User &\# Item &\# Edges &Split ratio\\
     \hline
     Gowalla &$29,858$ &$40,981$ &$1,027,370$ &$70/10/20$ \\
     Yelp2018 &$31,668$ &$38,048$ &$1,561,406$ &$70/10/20$ \\
     Amazon-Book &$52,643$ &$91,599$ &$2,984,108$ &$70/10/20$ \\
 \bottomrule
\end{tabular}
\label{dataset_stat_hete}
\vspace{-8pt}
\end{table}

\section{Model Details}
\label{sec:B}
In this section, we provide more details of the proposed PS2 methods from the neural architecture, hyper-parameter, and hardware perspectives. 

\subsection{Details of the Neural Architecture}
\label{sec:model}
Recall that our model consists of a personalized subgraph selector $g_\theta$, the GNN encoder $f_w$, and the link predictor $q_w$. The personalized subgraph selector is parameterized by a two-layer MLP with hidden dimension $D$ and output dimension $1$. The GNN encoder is a $K$-layer GCN~\cite{kipf2016semi} module, which varies from different downstream models. For example, when combining our PS2 with GAE and GraphSage, the default GNN module is GCN~\cite{kipf2016semi} and SAGE~\cite{hamilton2017inductive}. The link predictor is initialized as another three-layer MLPs. The hidden activation function in all neural networks is ReLU.

\begin{algorithm}[htb]
\vspace{-1pt}
   \caption{Alternating Gradient Descent for Eq.~\eqref{eq:bilevel}}
   \label{alg:example}
   \KwIn{Initial subgraph selector parameters $\theta$ and initial weight parameters $w$.}
   \While{not converge }{
   1. Upper-level optimization: Fix $w$, update subgraph selector $\theta$ by descending $\nabla_{\theta}\mathcal{L}_{val}(w-\lambda\nabla_{w}\mathcal{L}_{train}(w, \theta), {\theta})$.\\
   2. Lower-level optimization: Fix $\theta$, update weights parameters $w$ by descending $\nabla_{w}\mathcal{L}_{train}(w, \theta )$.}
\textbf{Return} Derive the optimal subgraphs for different edges based on the learned $\theta$ and $w$.
\vspace{-0pt}
\end{algorithm}

\subsection{Hyperparameter Configuration}
\label{sec:hyper}
To provide a fair comparison with state-of-the-art link prediction methods, we generally follow the same parameter settings across different baselines in terms of two different applications. In general, our model is optimized based on minibatch training. Following common practice for link prediction training, in each step, we sample a minibatch of positive edges from the training loader and then randomly generate one negative sample for each positive edge to construct the minibatch training set. Notice that we don't conduct subgraph sampling for node representation as done in~\cite{hamilton2017inductive}. We feed the whole adjacency matrix into the model for graph convolution. 
 
Specifically, for Planetoid datasets (Cora, CiteSeer, and PubMed), we adopt a three-layer GNN module with dimension 32. We set the batch size to 1024 and fixed the learning rate to 0.01. For OGB datasets, we adopt a three-layer GNN with the hidden dimension 256. The learning rate and batch size are fixed at 0.001 and 10 * 1024 as suggested in~\cite{hu2021ogb}~\footnote{\url{https://github.com/snap-stanford/ogb/tree/master/examples/linkproppred}}. For recommendation datasets, we adopt a three-layer GNN with the hidden dimension 64 according to~\cite{he2020lightgcn}. The batch size and learning rate are fixed as 1024 and 0.001, respectively. For different datasets, we search the hidden dimension $D$ of subgraph selector MLP layer from the set $\{64, 128, 256, 512, 1024\}$. The best options for three Planetoid and other datasets are 256 and 512, respectively. 

All the experiments are run 10 times, and we report the mean and the standard deviation.

\subsection{Hardware}
\label{sec:hardware}
We conduct all the experiments on a server with 48 Intel(R) Xeon(R) Silver 4116 CPU @ 2.10GHz processors, 188 GB memory, and four NVIDIA GeForce RTX 3090 GPUs.

\section{Gradient Approximation for Upper-level optimization}
\label{sec:C}
With $w$ fixed, the upper-level optimization updates $\bm \theta$ according to the validation performance as:
\begin{equation}
\begin{split}
    \theta^\prime = \theta - \lambda \nabla_\theta\mathcal{L}_{valid}(w^*(\theta), \theta).
\end{split}
\label{eq:upper22}
\end{equation}
However, evaluating the gradient \textit{w.r.t.} $\theta$ exactly is computationally prohibitive, since it requires solving for the optimal $w^*(\theta)$ whenever $\theta$ gets updated. To approximate the optimal solution $w^*(\theta)$, we propose to take one step of gradient descent update for $w$, without solving the lower-level optimization completely by training until convergence. 
Applying the chain rule, the approximated gradient yields:
\begin{equation}
\begin{split}
    \nabla_{\theta}\mathcal{L}_{valid}(w^\prime, \theta)-\lambda\nabla^2_{\theta,w}\mathcal{L}_{train}(w,\theta)\nabla_{{w}}\mathcal{L}_{valid}(w^\prime, \theta),
\end{split}
\label{eq:upper2}
\end{equation}
where $w^\prime=w-\lambda\nabla_{w}\mathcal{L}_{train}(w, \theta)$ is the weight for one-step forward model. The second term in Eq.~\eqref{eq:upper2} contains an expensive matrix-vector product, which requires $O(|\theta||w|)$ complexity. To further accelerate the optimization, we approximate the second term using the finite difference approximation, defined as:
\begin{equation}
\begin{split}
    &\nabla^2_{\theta,w}\mathcal{L}_{train}(w,\theta)\nabla_{w^\prime}\mathcal{L}_{valid}(w^\prime, \theta)
    \approx \frac{\nabla_\theta\mathcal{L}_{train}(w^+,\theta)-\nabla_\theta\mathcal{L}_{train}(w^-,\theta)}{2\epsilon},
\end{split}
\label{eq:upper3}
\end{equation}
Based on this approximation, we only need two forward passes for $w$ and two backward passes for $\theta$, therefore, the complexity is reduced from $O(|\theta||w|)$ to $O(|\theta| + |w|)$. The final result is 
\begin{equation}
    \begin{split}
        \nabla_\theta \mathcal{L}_{valid}(w^*(\theta), \theta)&\approx \nabla_\theta \mathcal{L}_{valid}(w^\prime, \theta) \\
        &- \lambda \frac{\nabla_{\theta}\mathcal{L}_{train}(w^+,\theta)-\nabla_{\theta}\mathcal{L}_{train}(w^-,\theta)}{2\epsilon},
    \end{split}
\label{weights_update22}
\end{equation}

\begin{table}[htbp]
\caption{Finetune vs. train from scratch on Planetoid dataset.}
  \vspace{-8pt}
 \begin{small}
\setlength{\tabcolsep}{2.5pt}
{
\begin{tabular}{l c c c}
\toprule
  &{Cora}
 &{CiteSeer}
 &{PubMed}\\
\midrule
GAE &$91.0 \pm 0.01 $ &$89.5\pm 0.04 $ &$96.4 \pm 0.01 $\\
GAE-PS2-scratch &$90.1\pm0.37$ &$90.2\pm0.9$ &$97.7\pm0.09 $\\
GAE-PS2 &$92.3\pm0.71$ &$92.2\pm0.19$ &$98.3\pm0.10$\\
\midrule
GraphSage &$86.3 \pm 1.06$ &$85.2\pm 2.56$ &$87.6\pm0.87$\\
GraphSage-PS2-scratch &$92.5\pm0.89 $ &$88.9\pm2.30$ &$90.9\pm0.44$\\
GraphSage-PS2 &$93.8\pm0.01$ &$93.4\pm1.20$ &$93.5\pm0.22$\\
\bottomrule
\end{tabular}}
 \end{small}
\label{table_lp_self}
\vspace{-8pt}
\end{table}

\begin{table}[ht]
\vspace{-0.5pt}
\centering
\caption{GNNLP vs. GNNLP-PS2 in terms of running time (in seconds) per epoch on three large-scale OGB datasets. }
\label{table_efficiency}
\begin{small}
\setlength{\tabcolsep}{1.5pt}
{
\begin{tabular}{lcccccc}
\toprule
 &ogbl-ddi &ogbl-collab &ogbl-ppa\\
\midrule
GAE:w.PS2 &$1.7:2.1 (\times 1.23)$  &$3.1:3.8 (\times 1.22)$ &$180.1:213.8 (\times 1.19)$\\
SEAL:w.PS2 &$282.1:285.4 (\times 1.01)$ &$36.4:40.7 (\times 1.12)$ &$1174.2:1397.9 (\times 1.19)$ \\

\bottomrule
\end{tabular}}
\end{small}
\vspace{-0pt}
\end{table}

\begin{figure}[ht]
  \centering
  \includegraphics[width=8.6 cm]{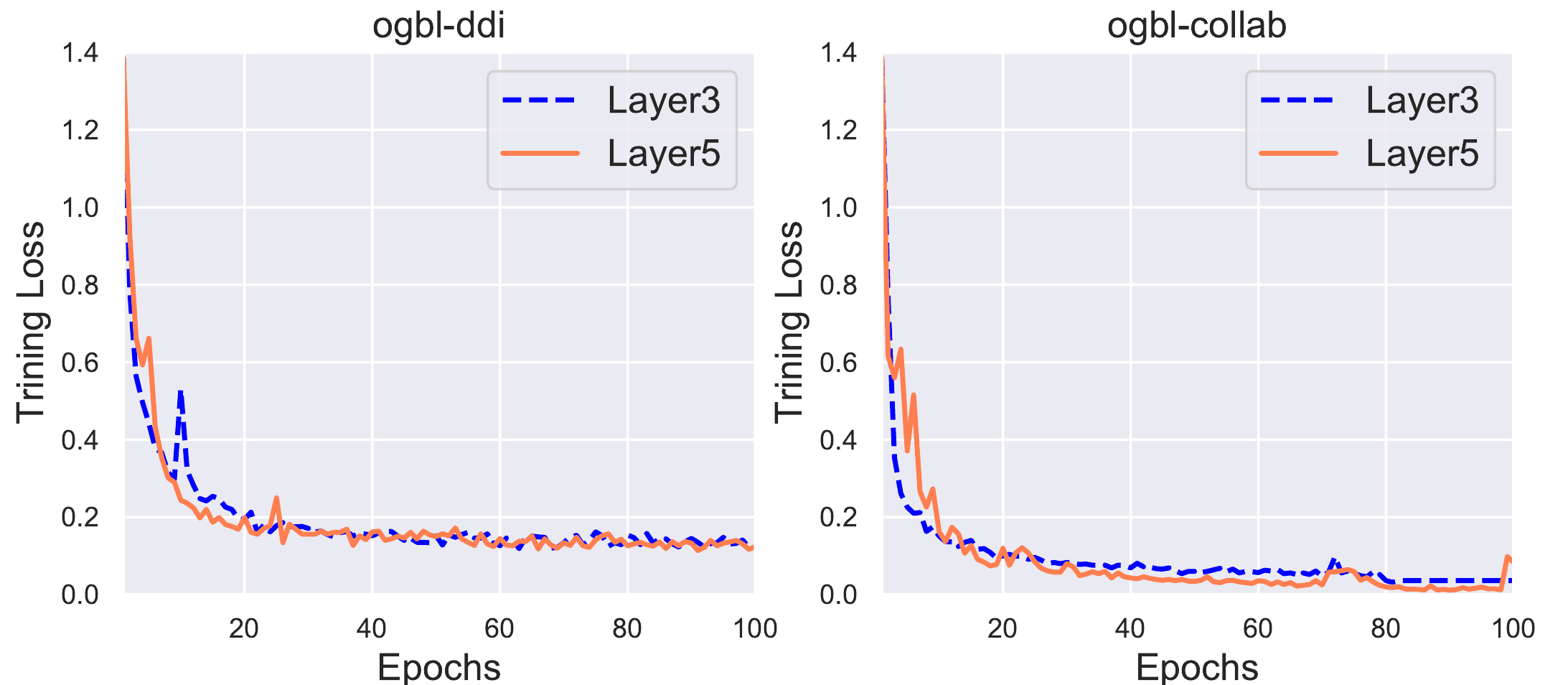}
  \caption{Empirical training curves of GAE-PS2 on datasets ogbL-ddi and ogbn-collab with different GCN~\cite{kipf2016semi} layers.
  }
  \label{train_losses}
\end{figure}

\end{document}